\documentclass[aps,pra,reprint,superscriptaddress,showpacs,showkeys]{revtex4-1}

\usepackage[utf8]{inputenc}
\usepackage{amsmath}
\usepackage{amssymb}
\usepackage{natbib}
\usepackage{graphicx}

\usepackage{textcomp}

\usepackage{mathtools}
\usepackage{upgreek}

\usepackage{xcolor}

\begin{document}

\title{Mutual coupling and synchronization of optically coupled quantum-dot micropillar lasers at ultra-low light levels}

\author{S\"oren~Kreinberg}
\affiliation{Institut f\"ur Festk\"orperphysik, Technische Universit\"at Berlin, Hardenbergstra{\ss}e 36, 10623 Berlin, Germany}

\author{Xavier Porte}
\thanks{ }
\affiliation{Institut f\"ur Festk\"orperphysik, Technische Universit\"at Berlin, Hardenbergstra{\ss}e 36, 10623 Berlin, Germany}

\author{David~Schicke}
\affiliation{Institut f\"ur Theoretische Physik, Technische Universit\"at Berlin,
	Hardenbergstra{\ss}e 36, 10623 Berlin, Germany}

\author{Benjamin~Lingnau}
\affiliation{Institut f\"ur Theoretische Physik, Technische Universit\"at Berlin,
	Hardenbergstra{\ss}e 36, 10623 Berlin, Germany}

\author{Christian~Schneider}
\affiliation{Technische Physik, Universit\"at W\"urzburg, Am Hubland, 97074 W\"urzburg, Germany}

\author{Sven H\"ofling}
\affiliation{Technische Physik, Universit\"at W\"urzburg, Am Hubland, 97074 W\"urzburg, Germany}
\affiliation{SUPA, School of Physics and Astronomy, University of St Andrews, St Andrews, KY16 9SS, United Kingdom}

\author{Kathy L\"udge}
\affiliation{Institut f\"ur Theoretische Physik, Technische Universit\"at Berlin,
	Hardenbergstra{\ss}e 36, 10623 Berlin, Germany}

\author{Stephan Reitzenstein}
\thanks{Corresponding authors: \mbox{stephan.reitzenstein@physik.tu-berlin.de,} javier.porte@tu-berlin.de}
\affiliation{Institut f\"ur Festk\"orperphysik, Technische Universit\"at Berlin, Hardenbergstra{\ss}e 36, 10623 Berlin, Germany}

\date{\today}

\begin{abstract}
In this work we explore the limits of synchronization of mutually coupled oscillators at the crossroads of classical and quantum physics. In order to address this uncovered regime of 
synchronization we apply electrically driven quantum dot micropillar lasers operating in the regime of cavity quantum electrodynamics. These high-$\upbeta$ microscale lasers feature cavity 
enhanced coupling of spontaneous emission and operate at output powers on the order of 100 nW. We selected pairs of micropillar lasers with almost identical optical properties in terms of the input-output dependence and the emission energy which we mutually couple over a distance of about 1\~m and bring into spectral resonance by precise temperature tuning. By  excitation power and detuning dependent studies we unambiguously identify synchronization of two mutually coupled high-$\upbeta$ microlasers via frequency locking associated with a sub-GHz locking range. A detailed analysis of the synchronization behavior includes theoretical modeling based on semi-classical stochastic rate equations and reveals striking differences from optical synchronization in the classical domain with negligible spontaneous emission noise and optical powers usually well above the mW range. In particular, we observe deviations from the classically expected locking slope and broadened locking boundaries which are successfully explained by the fact the quantum noise plays an important role in our cavity enhanced optical oscillators.  Beyond that, introducing additional self-feedback to the two mutually coupled microlasers allows us to realize zero-lag synchronization. Our work provides important insight into synchronization of optical oscillators at ultra-low light levels and has high potential to pave the way for future experiments in the quantum regime of synchronization.   

\end{abstract}


\flushbottom
\maketitle


\section{Introduction}
\label{section:introduction}
Synchronization is an ubiquitous phenomenon in mutually coupled systems~\cite{Pikovsky2003} which, under appropriate conditions, leads to a spontaneous self-organization of the coupled elements~\cite{ACE05}. 
A multitude of different physical, biological, or chemical systems can exhibit synchronization, making it a fundamental interdisciplinary property of interacting nonlinear systems~\cite{WIN80,Kuramoto.1984,PIK01}. 
The complexity  of this phenomenon is well depicted by the variety of existing synchronization scenarios, like e.g. chaos synchronization, where the individual coupled elements all follow the same chaotic trajectory~\cite{WIN90}. 
In that sense, semiconductor lasers coupled with a temporal delay represent an attractive and paradigmatic example of chaotic systems where synchronization has been fundamentally studied~\cite{Heil2001a,JAV03,ERZ06a,LIU11b} and widely explored for different regimes~\cite{Mirasso2004, Fischer2006, Ozaki2009, Tiana-Alsina2012, Aviad.2012, SOR13} and applications that range from random number generation to secure key exchange \cite{Kanter2010, Porte2016}.

Recently, the possibility of synchronization in nanoscale oscillators has raised increasing attention. 
Enabled by important technological advances, it has become feasible to explore nonlinear dynamics and synchronization at ultra-low energies in systems previously only explored from a quantum mechanical perspective. 
For instance, mutual synchronization of the Kuramoto type has been demonstrated in optomechanical structures~\cite{Heinrich2011} and in nanomechanical oscillators~\cite{Zhang2012, Matheny2014}. 
Most interesting is the quantum limit of nonlinear interaction, where novel phenomena have been predicted theoretically such as partial locking and synchronization blockade~\cite{Walter.2015, Lorch2017}.

\begin{figure}[hb]
	\includegraphics[width=0.95\columnwidth]{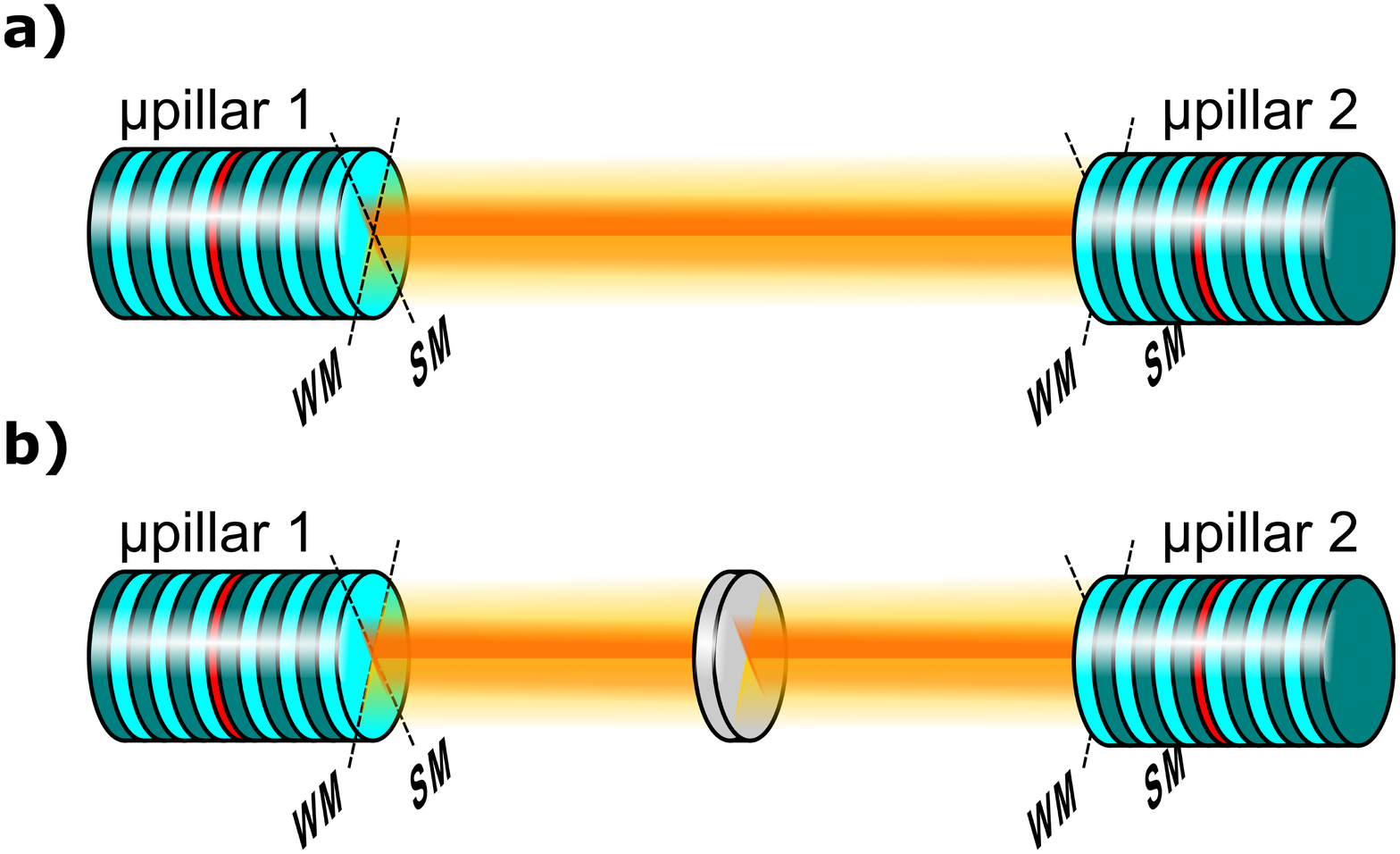}
	\caption{\label{fig:Scheme}Illustration of the studied experimental configurations. \textbf{(a)} Face-to-face mutual coupling and \textbf{(b)} mutual coupling via passive relay. The setup is arranged to couple the two perpendicularly polarized emission modes of each micropillar (i.e. strong mode (SM) and weak mode (WM)) to their respective counterparts.
	}
\end{figure}

Situated at the crossroads between nonlinear dynamics and quantum optics, cavity-enhanced microlasers offer a rich spectrum of exciting physics with potential applications as coherent light sources in system-on-chip quantum technologies~\cite{Munnelly2015}.
Due to their low mode volume on the order of the cubic wavelength, QD-microlasers operate in the regime where cavity quantum electrodynamics (cQED) effects become important. 
Up until now, microlaser studies have been focused almost exclusively on the properties of individual devices, not considering coupling interactions with external passive or active elements. 
However, this panorama is quickly changing and recent works report interesting effects like spontaneous symmetry-breaking due to local coupling between cavity modes in nanophotonic structures \cite{Hamel2015, Marconi2016}. 
Tailoring of the mode-switching dynamics and photon statistics have been shown in quantum dot (QD) microlasers subject to delayed optical feedback \cite{Albert.2011, Holzinger.2018}. 
Furthermore, first numerical works address the dynamics and stability of microlasers when mutually coupled with delay \cite{Han2016, Han2018}.


Here, we present the first experimental implementation of microlasers optically coupled with incoherent optical delay at $\ll~\upmu \mathrm{W}$ output power level in two different coupling configurations (cf. scheme in Fig. \ref{fig:Scheme}). 
We demonstrate mutual locking (see Fig. \ref{fig:Scheme}(a)) and synchronization (see Fig. \ref{fig:Scheme}(b)) via their spectral properties and photon statistics.
Furthermore, we explore the complex interaction between the bimodal emission of our micropillar lasers and its effect on the observed synchronization regimes. 
Accurate numerical modeling supports our findings and allow us to depict the time-resolved character of the synchronized dynamics. 

This paper is organized as follows. 
In Sec.~\ref{section:sampleandsetup}, we describe the microlaser samples and the experimental setup developed for coupling them. Section~\ref{sec:model} is devoted to present the numerical rate-equations model used to simulate the coupling behaviour of the microlasers. The results on the mutual locking and synchronization are presented and discussed in Sec.~\ref{sec:results}, before coming to the conclusions in Sec.~\ref{sec:conclusions}.


\section{Experimental Setup}
\label{section:sampleandsetup}
\begin{figure}[htb]
	\includegraphics{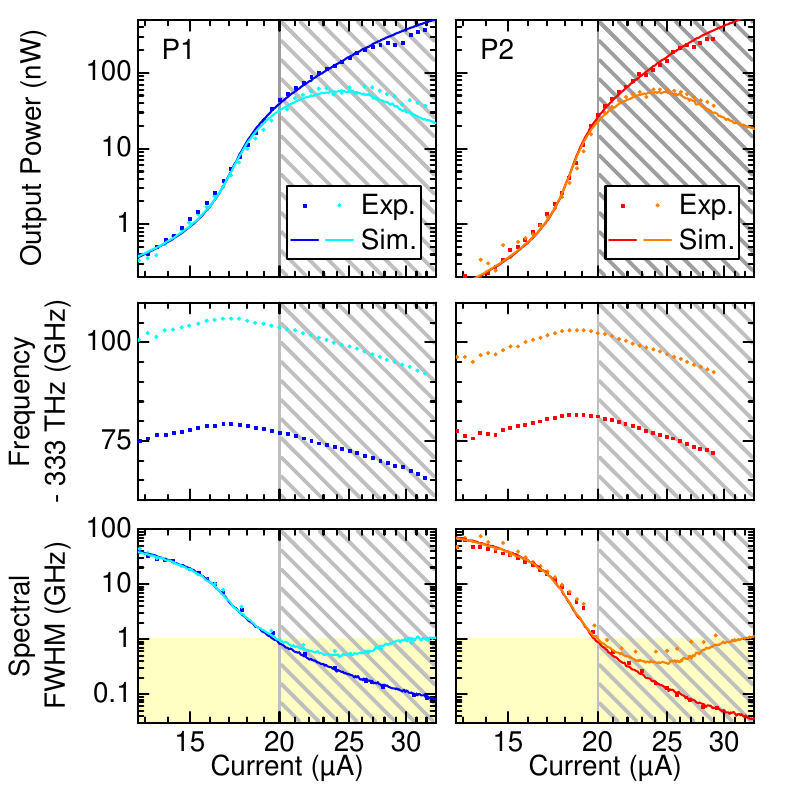}
	\caption{\label{fig:Input-Output}Experimental (dots) and numerical (lines) current dependence of output power, emission frequency and linewidth for the strong and weak modes of pillar 1 (left) and pillar 2 (right). Dashed areas indicate the operation regime for the mutual coupling experiments which require emission linewidths smaller than 1\,GHz (yellow area) and can be achieved for injection currents exceeding $20\,\mathrm{\upmu A}$. 
	}
\end{figure}

The nanolasers under study are $5~\upmu \mathrm{m}$ diameter electrically contacted micropillars based on AlGaAs heterostructures consisting of a single layer of In\textsubscript{0.3}Ga\textsubscript{0.7}As QDs with a density of $5\times10^9\,\mathrm{cm}^{-2}$ enclosed by two high-quality AlAs/GaAs distributed Bragg reflectors (DBR) (see SI for more technological details). 
This configuration ensures a small mode volume and pronounced light-matter interaction that result in cQED-enhanced coupling of spontaneous emission into the lasing mode~\cite{Reitzenstein.2008}. 
Such micropillars show high-$\upbeta$ lasing that is maintained by the gain contribution of a small number of QDs (cf. Sec.~\ref{sec:model}). 
The polarization degeneracy of the fundamental cavity mode is lifted by slight structural asymmetries resulting in two orthogonal linearly polarized components of the fundamental mode~\cite{Gayral.1998, Whittaker.2007, Reitzenstein2007}. 
The related bimodal behavior leads to a plethora of exciting physical phenomena such as gain competition~\cite{Leymann.2013}, unconventional  normal-mode coupling~\cite{Khanbekyan.2015}, and mode switching~\cite{Redlich.2016} as a instance of Bose-Einstein condensation of photons~\cite{Leymann.2017}. 
Moreover, it allows for the far-from-equilibrium generation of superthermal light~\cite{Marconi.2018}. 
Noteworthy, the competition for the common QD gain in micropillar cavities results in the divergence of the output powers and the linewidths of the split modes as soon as the lasing threshold is crossed (cf. top panels in Fig.~\ref{fig:Input-Output}). 
Thus, in the following we refer to the more intense mode as \emph{strong mode} and to the less intense one as \emph{weak mode}. 

Using advanced nanofabrication technology and an optimized sample design we realized dense arrays of 120 QD-micropillars each. 
Sample pieces each containing one of these arrays were placed into two independent He-flow cryostats separated by 700 mm and operated in a temperature range of 31\,K to 36\,K. 
The lasers in the two selected arrays come from neighboring parts of the same semiconductor wafer to ensure similar emission characteristics. The electrical current through the micropillar under investigation was estimated to be 1/120 of the current through the corresponding 120 micropillar array. 
For further details on the optical pre-characterization we refer to the supplementary information, section S-I.
Based on the detailed pre-characterization of all 240 microlasers which involved the measurement and evaluation of the output intensity, emission energy and emission linewidth of each microlaser as a function of the injection current, we selected the most suitable pair of microlasers which feature almost identical emission characteristics as presented in Fig.~\ref{fig:Input-Output}. 
The characteristic s-shape in the log-log input-output curves of the strong mode only exhibits a slight nonlinearity indicating a high $\upbeta$-factor. 
As previously mentioned, the above-threshold output power of the weak mode decreases with increasing pump current as a result of the intermodal gain competition~\cite{Redlich.2016}. 
Below threshold, the emission frequency of both modes shifts to higher values with increasing pump, while above threshold the frequency decreases with increasing pump because of sample heating. 
The frequency splitting of strong and weak mode stays fixed over the investigated pump current range. 
Due to the onset of coherence, the linewidth of the strong mode decreases more than two orders of magnitude eventually narrowing down to less than $100\,\mathrm{MHz}$. 
In contrast, the above-threshold weak mode linewidth increases, which indicates a non-complete transition to laser action for this mode. 
Previous injection-locking experiments in QD-micropillar lasers showed that a locking range of around 1\,GHz \cite{Schlottmann.2016} can be expected. 
In order to be able to resolve possible locking effects between the two microlasers, linewidths smaller than the expected locking range and hence output powers of at least 30\,nW are required (see Fig.~\ref{fig:Input-Output}, yellow and grey shaded areas respectively). 

\begin{figure}[htb]
	\includegraphics[width=0.98\columnwidth]{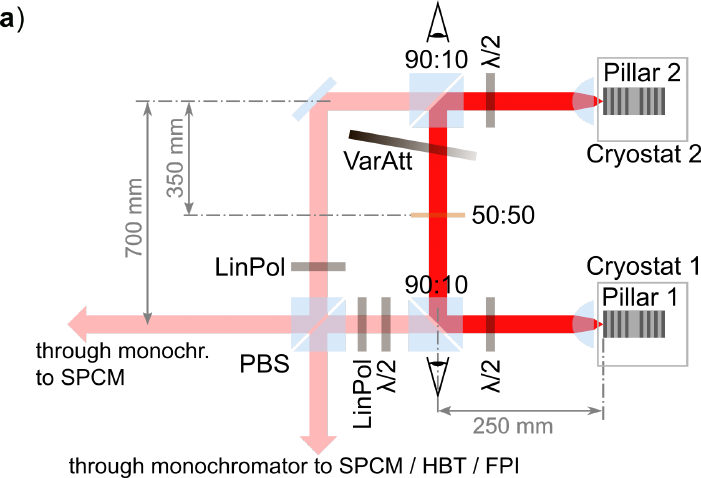}\\
	\vspace{0.5cm}
	\includegraphics{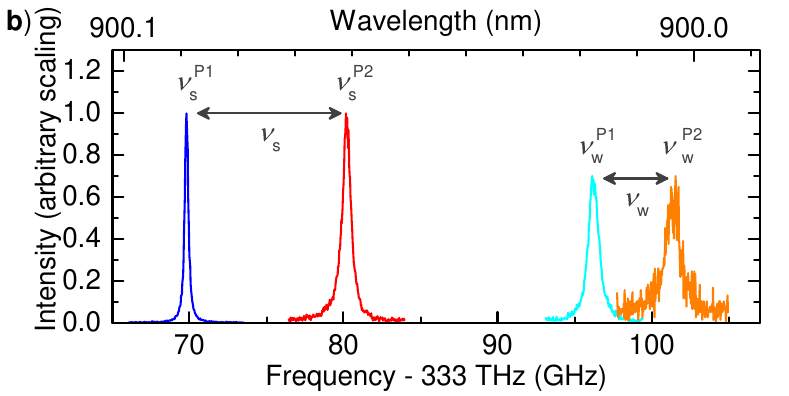}
	\caption{\label{fig:Setup-Spectra} \textbf{(a)} Experimental setup showing the coupling beam path (solid red) and the detection beam paths (pale red). 
	Each micropillar laser sample is placed in a cryostat at temperatures of $T_1 = 32\,\textrm{K}$ and $T_2 \in \left[32\,\textrm{K}, 36\,\textrm{K}\right]$. 
	Single-photon counting modules (SPCM) are used to measure autocorrelations in a Hanbury Brown and Twiss (HBT) configuration or crosscorrelations when placed after both monochromators.
	\textbf{(b)} Fabry-Perot interferometer (FPI) spectra of the non-coupled micropillar lasers. The strong mode and weak mode (respectively normalized to 1 and to 0.7) are depicted in the  
	same color-coding as used in Fig.~\ref{fig:Input-Output}. 
	}
\end{figure}

Fig.~\ref{fig:Setup-Spectra}(a) schematically shows the experimental setup which is used to study the mutual coupling of micropillar lasers via symmetric light paths. 
Emission of each microlaser is first collimated by an aspheric lense with significantly reduced transmission losses if compared to usually used long working-distance microscope objectives and is then directed by beam splitters with 90\% reflectivity to the other microlaser of the selected pair~\footnote{We would like to note that the use of an aspheric lens is crucial to achieve a high enough optical power level for the mutual coupling experiments between the microlasers.}. 
The transmitted light (10\%) is directed via a polarizing beam splitter (PBS) towards the two detection paths. 
Using polarization optics, it is possible to independently select the micropillar modes (strong and weak) being coupled and also those being detected. 
An optional variable attenuator (VarAtt) in the coupling path enables control of the coupling strength. 
In an ulterior version of the experiment which aims at the demonstration of zero-lag synchronization, the variable attenuator is substituted by a passive relay (pellicle mirror with 50\% transmittance) placed in the center of the beam path between both pillars. 
The required sub-micron mechanical stability of the coupling beam path between the pillars with microscale upper facets is ensured by a customized video control loop in which the microscopic image of each sample was constantly monitored by a computer such that (unavoidable) temperature-induced sample shifts were automatically compensated by tracking the motorized linear stages of the corresponding cryostat.

A sample spectrum of the emission modes of the selected pair of micropillar lasers is shown in Fig.~\ref{fig:Setup-Spectra}(b). 
The diagram illustrates the definition of the nominal detuning $\nu_\mathrm{s}:=\nu^{\mathrm{P2}}_\mathrm{s}-\nu^{\mathrm{P1}}_\mathrm{s}$ and $\nu_\mathrm{w}:=\nu^{\mathrm{P2}}_\mathrm{w}-\nu^{\mathrm{P1}}_\mathrm{w}$ of the strong and weak modes, respectively.
Due to a different frequency splitting between strong and weak modes in the two lasers, the strong mode and weak mode can be precisely and independently tuned in and out of resonance individually by sub-Kelvin temperature changes. 
We would like to note that the strong-weak mode splitting of each individual pillar depends mainly on structural asymmetries and was found to be independent of temperature and injection current. This can also be seen in Fig.~\ref{fig:Input-Output} (middle panels), showing the pump current dependence of the individual mode frequencies. In laser P1 (P2), a splitting of $26$~GHz ($21$~GHz) is found.
Consequently, the nominal detunings of weak modes and strong modes of the selected micropillar lasers differ by a constant value of 5\,GHz, $\nu_\mathrm{s}=\nu_\mathrm{w}+5\,\mathrm{GHz}$. 
By changing the injection current and thus, the output intensities of the two microlasers the coupling configuration can be continuously tuned from a master/slave scenario (where the output power of one laser is much larger and drives the weak laser) to a mutual coupling scenario (where the output powers of both lasers are similar) \cite{BON12}. 

In the following we first introduce the model to describe the mutual coupling of high-$\upbeta$ microlasers prior to presenting experimental data on the mutual coupling. 


\section{Model}
\label{sec:model}
\newcommand{\act}{{\rm act}}
\newcommand{\inact}{{\rm inact}}

The model used in this paper is based on semi-classical stochastic rate equations~\cite{Haken1989} taking into account the electron scattering mechanisms into the QDs as derived in our previous works \cite{Ludge2009,LIN15b, Redlich.2016}. The description of micro- and nanolasers with semi-classical equations was recently shown to be valid down to a surprisingly low number of emitters on the order of 10~\cite{MOR18}. Our chosen theoretical framework should therefore be suited to accurately describe the dynamical properties of the micropillar lasers considered here. In our model we account for the two orthogonal linearly polarized micropillar modes by two separate complex electric field equations, denoted as weak mode and strong mode, corresponding to their respective output power above threshold as discussed in the previous section. As the microlaser output is predominantly linearly polarized and dominated by strong spontaneous emission, we couple both laser modes to a single charge-carrier type and describe the mode interaction by phenomenological gain compression terms. We thus neglect spin-flip dynamics required to model the dynamics in lower-$\upbeta$ VCSEL devices \cite{SAN06c,VIR12}.
For each of the two coupled lasers, we model the electrical fields $E_j$ of the two modes $j\in{\mathrm{w},\mathrm{s}}$, the occupation probability of the active and inactive quantum dots $\rho_{\rm (in)act}$, and the reservoir carrier density $n_r$. Here we denote as active the portion of QDs within the inhomogeneous distribution that couple to the lasing mode via stimulated emission. 

\begin{eqnarray}
\label{modelE} \nonumber \frac{{\rm d}}{{\rm d} t}E_j(t) &=& \left[\tfrac{\hbar\omega Z^{QD}}{\epsilon_0\epsilon_{bg} V} g_j
\big(2\rho_\act(t)-1\big) - \kappa_j\right](1+i\alpha) E_j(t) + \\
&&\frac{\partial}{\partial t}E_j \big|_{sp} + \frac{\partial}{\partial t}E_j \big|_{coup} \\
\label{modelrho} \nonumber \frac{{\rm d}}{{\rm d} t}\rho_\act(t) &=& -\sum_{\mathclap{j\in\{s,w\}}} g_j\big[2\rho_\act(t)-1\big]
|E_j(t)|^2- \frac{\rho_\act(t)^2}{\tau_{sp}} + \\
&&S^{in} n_r(t) \big[1-\rho_\act(t)\big] \\
\label{modeln} \nonumber \frac{{\rm d}}{{\rm d} t} n_r(t) &=& \tfrac{\eta}{e_0 A}(J-J_{p})  - S^{in} n_r(t)  \tfrac{2 Z^{QD}}{A}\big[1-\rho(t)\big] \\
&&-\frac{n_r(t)}{\tau_{r}} - \frac{2Z^{QD}_\inact\rho_\inact}{A\tau_{sp}}
\end{eqnarray}

The laser is pumped by injecting an electric current $J$ into the reservoir $n_r$ from where electrons may either recombine without contributing to the lasing mode or scatter into QDs with the rate $S^{in}\times n_r(t)$. We account for experimental details in the pumping process by assuming an injection efficiency $\eta$, and a parasitic current $J_p$, determined from fits to the experimental input-output curves, see also Fig.~\ref{fig:Input-Output} and Table $\ref{table}$. The occupation of inactive dots is calculated from the steady-state value of Eq.~\eqref{modelrho} without stimulated emission, taking into account only spontaneous recombination within these dots:
\begin{align}
  \label{modelrhoIA}\rho_\inact(t) &= (\tau_{sp}S^{in}n_r)(1+\tau_{sp}S^{in}n_r)^{-1}
\end{align}

The electric fields of weak mode and strong mode both interact with the active QDs by stimulated emission. Since the frequencies of the two modes differ by only a few tens of $\upmu$eV, we consider only one carrier population that is interacting with both optical modes, which leads to gain competition, modeled as 
\begin{equation} 
\label{modelg} g_j = g_j^0 \Big(1+\varepsilon_0 n_{bg} c_0 \sum_{i\in\{s,w\}}\varepsilon_{ji} |E_i(t)|^2 \Big)^{-1}
\end{equation}
The gain $g_{\mathrm{s},\mathrm{w}}$ of strong and weak modes respectively depends on the individual intensity of both modes and the compression factors $\varepsilon_{ij}$ with $i,j\in \{\mathrm{w},\mathrm{s}\}$. 
A mode with high intensity reduces (compresses) the gain for both modes. 

Spontaneous emission into the lasing modes is modeled via a Gaussian white noise source $\xi(t)\in\mathbb{C}$, where $\langle \xi(t) \rangle =0$ and $\langle \xi(t) \xi(t') \rangle= \delta (t-t')$, such that
\begin{align}
\frac{\partial}{\partial t}E_j \big|_{sp}=\sqrt{\upbeta \frac{\hbar \omega}{\epsilon_0 \epsilon_{bg}} \frac{2 Z^{QD}}{V} \frac{\rho_\act^2}{\tau_{sp}} } \xi(t). \label{eq:sponE}
\end{align}

We simulate the two coupled micropillar lasers each with its own set of differential equations \eqref{modelE}-\eqref{modeln}, with the two lasers indicated by an index $\mathrm{P1},\mathrm{P2}$, respectively. In the rotating frame of the free-running emission frequency of P2, the mutual coupling of the two lasers is expressed by 
\begin{eqnarray} 
\nonumber \frac{\partial}{\partial t}E_j^{\mathrm{P1}} \big|_{coup} &=& K\kappa^{\mathrm{P1}}_j E^{\mathrm{P2}}_j(t-\tau) 
  + 2\pi i \nu_j E_j^{\mathrm{P1}}, \\
\nonumber \frac{\partial}{\partial t}E_j^{\mathrm{P2}} \big|_{coup} &=& K\kappa^{\mathrm{P2}}_j E^{\mathrm{P1}}_j(t-\tau),
\end{eqnarray}
where $K$ is the coupling strength and $\tau$ the time delay after which the light from one laser arrives at the other. 
The term $\nu_s$ accounts for the relative frequency detuning between the two strong modes, with an additional 5 GHz detuning between the weak modes due to the mode splitting mentioned above: $\nu_\mathrm{w} = \nu_\mathrm{s} + 5\,{\rm{GHz}}$.

Using the above model, we can accurately reproduce the measured input-output characteristics and current-dependent linewidths (see lines in Fig.~\ref{fig:Setup-Spectra}), and allows for an accurate extraction of model parameters from the measured data. The slight differences in the laser characteristics between the two microlasers lead also  to slightly different input parameters for the modeled devices. The parameters used in the simulations are listed in Table $\ref{table}$.

\begin{table}
	\centering
	\begin{ruledtabular}
		\begin{tabular*}{\textwidth}{@{\extracolsep{\fill} }lrl}
			Fitted Parameter &  & Value\\
			    \hline
			Optical cavity losses, & &\\
			strong (weak) mode & $\kappa_\mathrm{s}$ ($\kappa_\mathrm{w}$) & $39~(38.5)$  \,ns$^{-1}$ \\
                        Optical gain coefficient, & &\\
                        strong (weak) mode & $g^0_\mathrm{s}$ ($g^0_\mathrm{w}$) & $5.35~(5.21)$  \, $\frac{{\rm m}^2}{{\rm V}^2{\rm s}}$ \\ 
			Self gain compression, & &\\ 
			strong (weak) mode & $\varepsilon_{\mathrm{ss}}$ ($\varepsilon_{\mathrm{ww}}) $& $10~(12)~\cdot 10^{-10} \frac{m^2}{W}$ \\ 
			Cross gain compression, & &\\ 
			strong (weak) mode & $\varepsilon_{\mathrm{sw}}$ ($\varepsilon_{\mathrm{ws}}$) &$16~(17.8)~\cdot10^{-10} \frac{m^2}{W}$ \\ 
			Spontaneous emission factor & $\upbeta^{\mathrm{P1}(\mathrm{P2})}$ & $3.5$ $(4)\, \cdot \,10^{-3}$ \\
			Parasitic current & $J_{p}^{\mathrm{P1}(\mathrm{P2})}$&  $2.3$ $(7.3)$\,$\upmu$A\\
			Pump efficiency & $\eta^{\mathrm{P1}(\mathrm{P2})}$ & $0.596$ $(0.674) $ \\
			Linewidth enhancement factor & $\alpha^{\mathrm{P1}(\mathrm{P2})}$ & $1.7$ $(1.0)$\\
			Reservoir carrier lifetime & $\tau_{r}$ & $1$\,ns\\
			    \toprule
			Given Parameter &  & Value\\
			    \hline
			Effective scattering rate & $S^{in}$ & $7\cdot 10^{-15}$\,m$^2$\,ps$^{-1}$\\
			Effective lasing mode area & $A$ & $15$\,µm$^2$ \\
			Lasing mode volume & $V$ & $5$\,µm$^3$\\ 
			Number of (in)active QDs & $Z^{QD}_{(\inact)}$ & $312$ $(938)$ \\
			Background refractive index & $n_{bg}$ & $3.34$\\
			QD lifetime, $\upmu$-laser 1 (2) & $\tau_{sp}^{\mathrm{P1}(\mathrm{P2})}$ & $155$ $(185)$\,ps \\
			Photon energy & $\hbar\omega$ & $1.38$\,eV \\
			Coupling delay time & $\tau$ & $3.85$\,ns\\
		\end{tabular*}
	\end{ruledtabular}
	\caption{Parameters used for the simulations if not stated
		otherwise. 
	} \label{table}
\end{table}


\section{Results}
\label{sec:results}

In the following we characterize the synchronization of our mutually coupled micropillar lasers via their spectral and photon-statistic properties. 
We first study the spectral properties, unveiling a coherence behavior and locking properties particular to high-$\upbeta$ microlasers. 
We use second-oder correlation functions to experimentally observe different types of synchronization in our mutually coupled micropillar lasers, which are backed up by real-time dynamics from numerical simulations.
Furthermore, the combination of spectral and correlation analyses turns out to be a powerful toolbox to understand the complex mode-interactions in cavity-enhanced bimodal microlasers.
 

\subsection*{Locking range width versus coupling strength}
\begin{figure}
	\includegraphics[width=0.9\columnwidth]{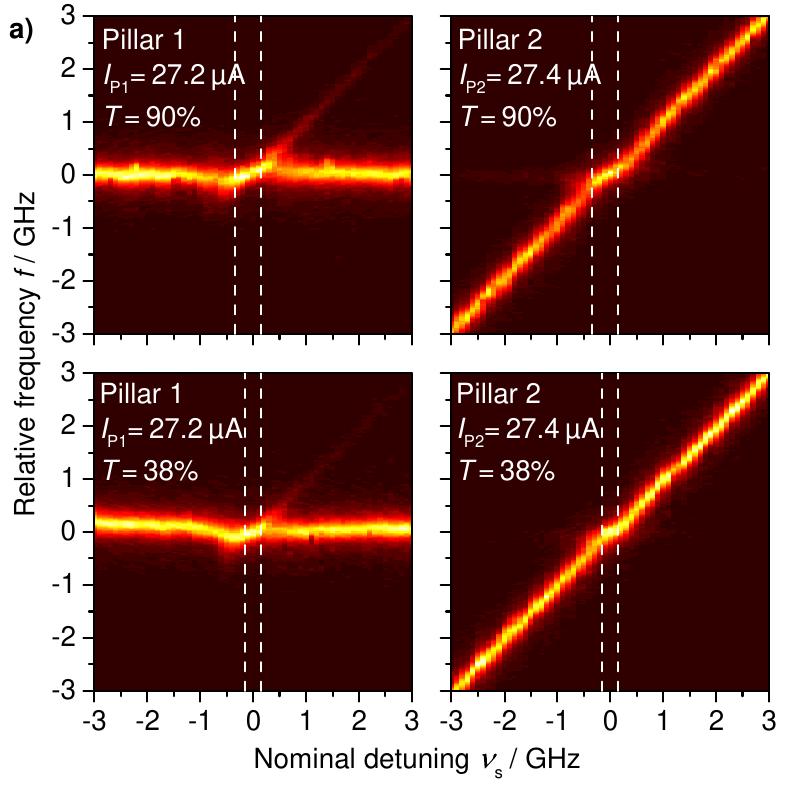}\\
	\includegraphics[width=0.9\columnwidth]{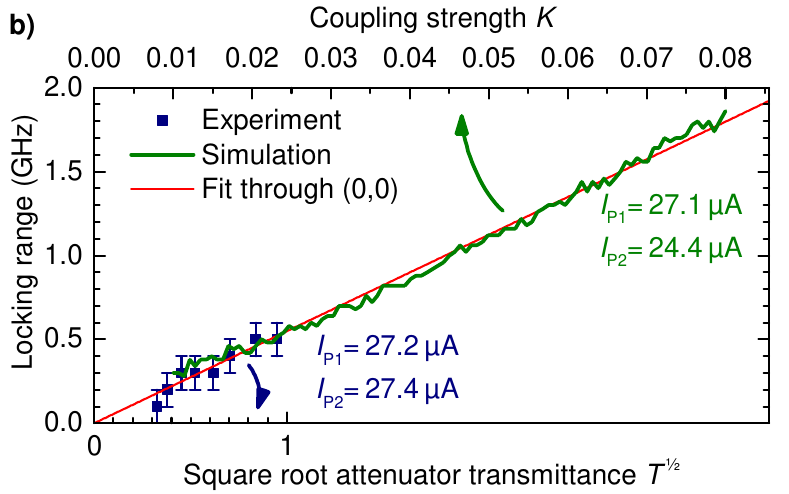}
	\caption{\label{fig:Attenuation-of-mutual}Mutually coupled strong modes of the two micropillar lasers for different coupling strengths K.
	\textbf{(a)} Detuning scans of the strong modes with high (upper panels) and low (lower panels) coupling strengths. 
	$T$ is the transmittance of the attenuator in the coupling path. 
	\textbf{(b)} 
	Dependence of the locking range on the square root of the attenuator transmission (lower axis) and the numerical simulation coupling strength $K$ (upper axis). 
	The horizontal axes are scaled such that the displayed linear fits to experiment and simulations are superimposed. 
	}
\end{figure}

In a central experiment of this paper we investigate the locking properties of the selected pair of mutually coupled micropillar lasers. 
For this purpose we vary the relative detuning between the two coupled microlasers as shown in Fig.~\ref{fig:Attenuation-of-mutual}(a). 
The emission frequency of pillar 1 is kept constant (at constant temperature of 32\,K), meanwhile the frequency of pillar 2 is precisely scanned across the emission frequency of pillar 1 by sweeping its temperature in the range $T_2 \in \left[ 32\,\textrm{K},  36\,\textrm{K}\right]$. 
While the temperature is swept, emission spectra of pillar 1 are recorded by using the Fabry-Perot scanning interferometer. A matrix is formed from the spectra, such that each column of the matrix corresponds to one spectrum. Emission spectra of pillar 2 are recorded in the same way in a second run. The matrices are then plotted as 2D heat maps. The detuning ranges of $\pm 3~\mathrm{GHz}$ displayed in Fig.~\ref{fig:Attenuation-of-mutual}(a) correspond to a temperature range of 34.9\,K to 33.9\,K.
When tuning the two lasers close to resonance, i.e. for detunings $\lesssim 0.5$~GHz, clear mutual frequency locking can be identified as a change in slope of the relative frequency vs. detuning characteristics: Within the locking range, the emission of both lasers is shifted towards a common frequency, returning to their free-running values outside of the locking range. 
A comparison between upper and lower panels of Fig.~\ref{fig:Attenuation-of-mutual}(a) illustrates that the locking range depends on the mutual coupling strength (varied by adjusting the variable attenuator in the coupling path), which has a transmittance $T$ of 90\% (38\%) in the upper (lower) panel. 

Deeper insight into the locking behavior requires a more detailed study of the locking range as a function of the coupling strengths. 
In agreement with previous reports on externally controlled micropillar lasers \cite{Schlottmann.2016} and coupled semiconductor lasers \cite{WUE05a,VIC06}, the locking range width is proportional to the injected electric field strength, i.e. to the square root of the attenuator transmittance (T). Figure ~\ref{fig:Attenuation-of-mutual}(b) depicts such dependence for both experimental (symbols) and simulations (solid line) data.
In order to plot together both data, we match the linear dependence of the locking range on $K$ (numerics) with the experimental data to find the proportionality factor between $K$ and the square root attenuator transmittance \footnote{This matching is necessary because while a measurement of the free space optical losses (beam splitters, polarization optics, lenses and cryostat windows) is in principle possible, it is not possible to quantify the coupling efficiency into the pillar and to the laser field.}. The maximum experimental amplitude coupling strength ($\mathrm{T}=1$) is thus estimated as $K \approx 2.5\%$. In the simulations, the coupling strength $K$ is studied over a larger range. 

\subsection*{Study of the mutual coupling}

\begin{figure}
        \includegraphics{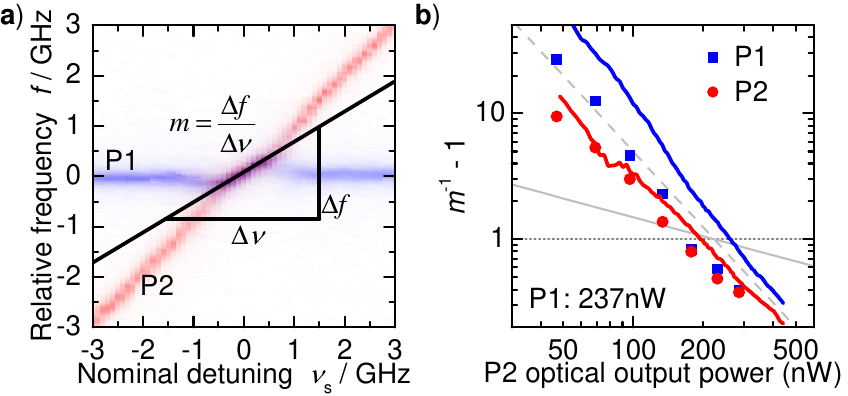}
        \caption{\label{fig:Locking-slope-in}Locking slopes of the two mutually coupled QD-microlasers vs. output power of pillar P2.
        Panel \textbf{(a)} illustrates how the slope $m$ is calculated. \textbf{(b)} Experimental (symbols) and numerically simulated (lines) locking slopes in dependence of the optical output power of pillar P2. The horizontal dotted line depicts the classically expected slope $m = 0.5$ and the oblique dashed and continuous grey lines respectively correspond to the slopes of $A = -2$ and $A = -0.5$. 
        }
\end{figure}

The presence of locking between the microlasers emission unequivocally indicates coupling. However, it is the slope $m$ described by the microlasers' emissions inside the locking region (see Fig.~\ref{fig:Attenuation-of-mutual}(a)), which determines the direction of the coupling. Figure~\ref{fig:Locking-slope-in}(a) depicts this slope as the ratio of frequency change of the locked signal $\Delta f$ and the nominal detuning $\Delta\nu$. We use this slope as the indicator for having achieved not only unidirectional but mutual coupling. 
Consider for instance the limiting case of a unidirectional injection experiment: Here the emission of the injecting master laser by definition must not be influenced by the slave laser subjected to injection. 
Strictly speaking this condition can only be fulfilled by placing optical isolators in the coupling path. However, even without an optical isolator \textendash{} if only the output powers of the mutually coupled lasers are strongly imbalanced \textendash{} there will be one \textquotedblleft master-like'' laser and a \textquotedblleft slave-like'' laser. While the former is almost unaffected by the mutual coupling, the latter is strongly influenced by the injected light. In this situation, when tuning the master laser, the slave-laser will perfectly follow the injected signal in the locking region, which results in a locking slope of $m = 1$. On the contrary, if only the slave-like laser is tuned, the locking slope will have a value of $0$, because its emission frequency is locked to the master-like laser. If the output power imbalance between master and slave is reduced, the locking slope will start turning away from these extreme values and eventually reach $m=0.5$ for evenly balanced coupling (cf. horizontal dotted line in Fig.~\ref{fig:Locking-slope-in}(b)). 

Based on these considerations, for phase-locked (or frequency-locked) lasers under mutual coupling conditions, the locking slopes of both oscillators, $m_{\mathrm{P1}}$ and $m_{\mathrm{P2}}$, would be expected to be equal, as both lasers are locked to each other and emit light on a frequency in between the two free-running laser lines. Surprisingly, both in experiment and simulations, the two microlasers exhibit different locking slopes. In Fig.~\ref{fig:Locking-slope-in}(b), the slopes $m_{\mathrm{P1}}$ and $m_{\mathrm{P2}}$ can be seen to differ especially for low output powers of pillar 2, which resembles a master-slave setup for which $m_{\mathrm{P2}}=m_{\mathrm{P1}}=0$ is expected. 
This means that inside the locking range, the average emission frequency of the two microlasers is deviating proportionally to the nominal detuning. These deviations are attributed to the effect of partial locking in high-$\upbeta$ microlasers~\cite{Schlottmann.2016,Cresser2.1982}.  The fact that the locking slopes get more similar when the output power of pillar 2 is increased, is explained not only by the stronger injection into pillar 1 but also by the decreasing relative contribution of quantum noise to the output power of pillar 2. 

To theoretically analyze our experimental and numerical observations, we reduce our laser model to a system of coupled phase oscillators \cite{Kuramoto.1984}. We do so by neglecting the amplitude dynamics of the electric fields within the microlasers and setting the linewidth enhancement factor $\alpha=0$. 
The resulting phase equations read
\begin{align*}
\dot{\varphi}_{1}(t) & =\varepsilon_{2\to1}\sin(\varphi_{2}(t-\tau)-\varphi_{1}(t))\\
\dot{\varphi}_{2}(t) & =\varepsilon_{1\to2}\sin(\varphi_{1}(t-\tau)-\varphi_{2}(t)) + 2\pi \nu
\end{align*}
In order to quantify the locking dynamics, we define the locking slope $m$,
\begin{align*}
 m := \frac{{\rm d}f}{{\rm d}\nu}\,,
\end{align*}
where $2\pi f := \dot \varphi_1 = \dot \varphi_2$ is the common phase velocity of the mutually locked oscillators. A locking slope of $m=0$ or $m=1$ denotes the limit cases where the locked oscillation frequency of both oscillators is given by the free-running frequency of oscillator $1$ or $2$, respectively.

Within this approach, the locking slope $m$ depends on the quotient of the coupling-strengths,
$$\varepsilon_{n\to m} = K \kappa_j^{{\rm P}m} \left|\frac{E_j^{{\rm P}n}}{E_j^{{\rm P}m}}\right|\,,$$
and can be calculated approximately to:
\[
{m_{\mathrm{P2}}}^{-1}-1 \approx \frac{\varepsilon_{1\to2}}{\varepsilon_{2\to1}} + 2 \tau\, {\varepsilon_{1\to2}} {\textstyle .}
\]
The second term on the r$.$h$.$s$.$ dominates the locking slope for all cases considered here. 
Hence, for fixed output power of pillar P1, ${m_{\mathrm{P2}}}$ depends on the output power $P_{out,2}$ of pillar P2:
\[
(P_{out,2})^{-\tfrac{1}{2}}\propto {m_{\mathrm{P2}}}^{-1}-1
\]
With increasing power of pillar P2, the common frequency within the locking range is thus pulled closer towards the free-running frequency of P2. Therefore we fit the equation
\begin{align}
  B\cdot (P_{out,2})^{A}={m_{\mathrm{P2}}}^{-1}-1 \label{eq:slopemodel}
\end{align}
to the experimental data. 
In contrast to the analytic expectations, the experimental data and numerical simulations suggest an exponent of $A \approx -2$ instead of the expected $A=-\tfrac{1}{2}$ (see respective dashed and continuous grey lines in Fig.~\ref{fig:Locking-slope-in}(b)). This behavior will be further investigated in the following section.

\begin{figure}
        \includegraphics{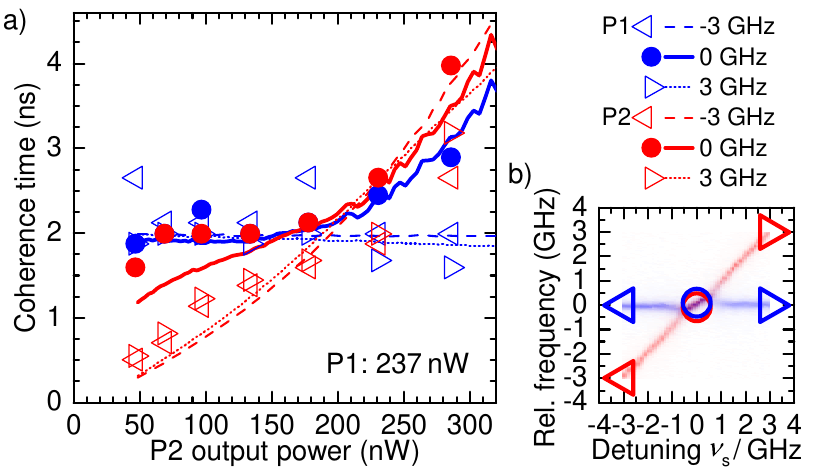}      
        \caption{\label{fig:Coherence-time-of}\textbf{(a)}~Coherence times of the mutually coupled micropillar lasers
        vs. output power of laser P2 (blue: pillar 1, red: pillar 2). The coherence times are determined from the optical linewidths. \textbf{(b)}~Exemplary locking diagram to illustrate the detunings within ($\nu_s=0$, circles) and outside the locking range ($\nu_\mathrm{s} = \pm3\,{\rm GHz}$ triangles) from where the coherence times are calculated. In addition the solid and dashed/dotted lines in panel (a) show the simulation results for the laser coherence time inside and outside of the locking range, respectively. 
                }
\end{figure}

\begin{figure}
        \includegraphics{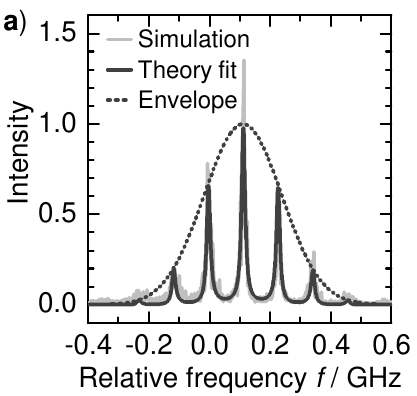}\hspace*{\fill}\includegraphics{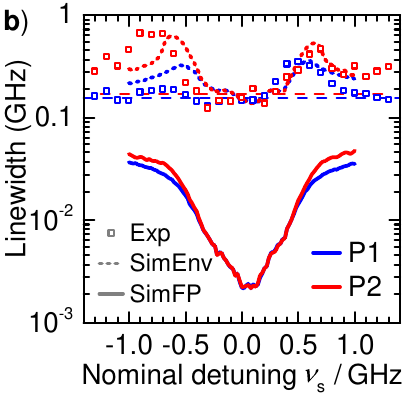}
        \caption{\label{fig:simulated_spectra_and_linewidth}Emission spectra and linewidths of the mutually coupled micropillar lasers for pump currents of $I_{\mathrm{P1}}=27.0\,\mathrm{\upmu A}$ and $I_{\mathrm{P2}}=24.0\,\mathrm{\upmu A}$. 
        \textbf{(a)} Simulated spectrum of pillar P1 in a mutual coupling setup (light gray), along with a fit to the spectrum (dark gray) using a Fabry-Perot interferometer-like spectral transmission function with a Gaussian envelope (dashed gray). \textbf{(b)} Extracted linewidths of the Gaussian envelope (dotted line) and individual Fabry-Perot lines (solid line) for pillar P1 (blue) and P2 (red). As reference, the free-running linewidth is shown in dashed lines. Linewidths from fits to experimental spectra for $I_{\mathrm{P1}}=28.8\,\mathrm{\upmu A}$ and $I_{\mathrm{P2}}=24.6\,\mathrm{\upmu A}$ are plotted as open squares.
    }
\end{figure}

\subsection*{Coupling enhancement of the coherence times}


The optical coupling of semiconductor lasers was previously shown to improve the coherence properties of both lasers by suppressing the noise-induced phase drift \cite{AGR84,HEG07,BRU17,KEL17a}. The high $\upbeta$-factor of the microlasers makes them a strongly noise-dominated system. Studying the dependence of the noise and coherence properties of the coupled microlasers depending on the control parameters is thus an important aspect of the cavity-enhanced mutually coupled oscillators. To explore the underlying physics, we extract the coherence times of the coupled microlasers from their spectral linewidth, both in the locked region for small detuning and for unlocked lasers with a detuning of $\nu_\mathrm{s}=\pm3$\,GHz, as shown in Fig.~\ref{fig:Coherence-time-of}. 
We evaluate the coherence time in dependence of the output power of pillar P2, while keeping the voltage of pillar P1 constant. 
This way, the coupling scheme can be tuned between master-slave-like coupling (low P2 power) and symmetric mutual coupling (equal power) of P1 and P2. 
The coherence times outside of the locking ranges can be seen to be only weakly influenced by the coupling, and just increase with increasing laser power. The coherence time of laser P1 stays constant outside of the locking range, as its output power stays constant. This indicates that a small detuning of 3~GHz between P1 and P2 only weakly influences the coherence properties of both lasers. 
Within the locking range (circles in Fig.~\ref{fig:Coherence-time-of}), when the output power of pillar P2 is below that of pillar P1 ($237\,\mathrm{nW}$), a pronounced improvement of the coherence time of P2 towards that of P1 can be observed in the locking range. In contrast, for higher output power of P2, the coherence time is pulled towards that of P2. 
The coherence time of the mutually locked lasers is therefore determined predominantly by the stronger laser \cite{MAL65}, which is also the laser with higher coherence time.

Interestingly, the numerical simulations reveal additional spectral features within laser line. Figure~\ref{fig:simulated_spectra_and_linewidth}(a) illustrates the presence of a fine structure in the emission spectra. The spectra are composed of a regular frequency comb with $\approx130$\,MHz spacing, corresponding to the total round-trip coupling delay of $2\tau=7.7$\,ns. We interpret the resulting spectral shape as a stochastic excitation of different compound laser modes (CLMs), i.e., standing waves within the combined cavity formed by the coupled micropillar lasers \cite{ERZ06a}. The stochastic switching between different CLMs leads to the presence of many different spectral peaks, weighted with a Gaussian envelope function \cite{DHU14}.
Experimentally, this fine structure cannot be resolved due to insufficient spectral resolution of the FPI. For comparison with the experiment, the resulting numerical spectra must be convolved with an artificial Lorentzian-detector response function. 
Figure~\ref{fig:simulated_spectra_and_linewidth}(b) depicts the experimental and numerical (both raw and convolved) linewidth dependences with respect to the nominal detuning between the two lasers. 
While the spectral width of the Gaussian envelope (dotted lines), which is in good agreement with the experimental data (open squares), is reduced only down to the free-running laser linewidth, the individual Fabry-Perot modes (solid lines) exhibit a strong narrowing inside the locking range, with linewidths down to a few MHz. This indicates strong coherence within each of the compound laser modes.
At the locking boundaries, the width of the Gaussian envelope is observed to exceed the free-running laser linewidth. This is a signature of dynamical instabilities at the locking boundaries, leading to a strongly reduced coherence of the laser light output near the unlocking transition. The underlying bifurcation structure of the deterministic system is strongly washed out due to the noise-dominated nature of the high-$\upbeta$ microlasers. In this highly stochastic regime, we therefore rely on spectral and correlation properties to more comprehensively characterize the laser dynamics.


\subsection*{Intensity auto- and cross-correlations}

In the field of cavity-enhanced nano- and microlasers a detailed study of the photon statistics of emission is of particular interest. 
Measuring the power-dependent photon autocorrelation function $g^{(2)}(\tau)$ allows for instance for the unambiguous proof of laser emission in high-$\upbeta$ lasers operating close to the limit of the thresholdless regime~\cite{Ota.2017}, for the identification of superradiant emission~\cite{Jahnke.2016}, or for ruling out chaotic mode switching~\cite{Schlottmann.2016}. Additionally, it is also highly beneficial for the identification of chaotic dynamics in feedback coupled microlasers operating at ultralow-emission powers~\cite{Albert.2011}. 

\begin{figure}
	\includegraphics{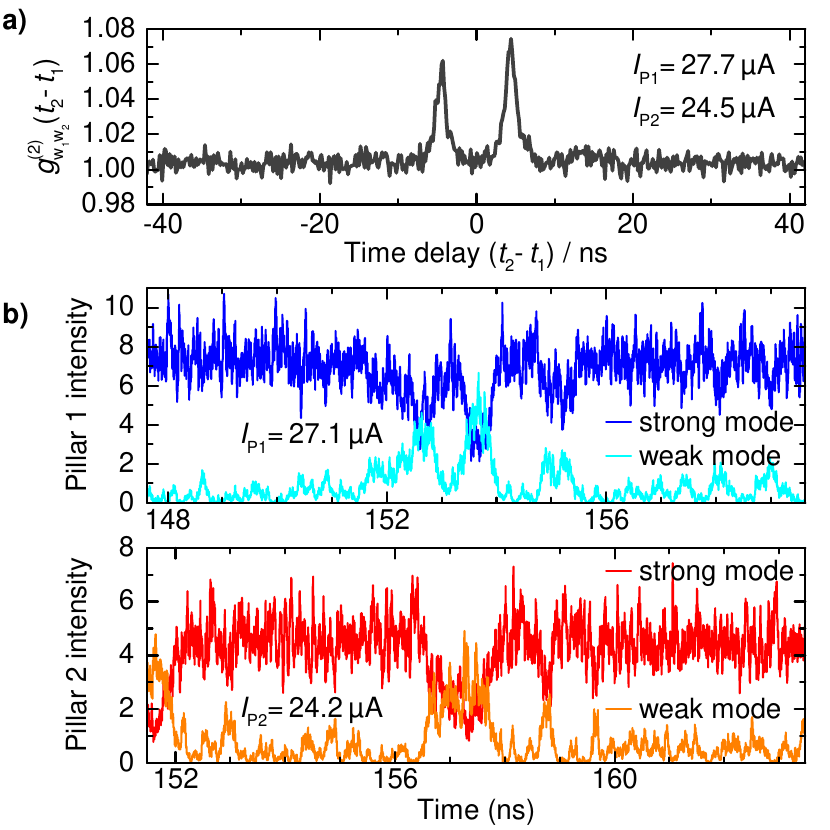}
	\caption{\label{fig:Intensity-crosscorrelation-of}\textbf{(a)} Intensity cross-correlation $g_{w_1w_2}^{(2)}(t_2-t_1)$ of the weak modes of the pillars
	with the weak modes tuned to resonance ($\nu_\mathrm{w}=0$). 
	The two main peaks at \textpm 4.3\,ns suggest leader-laggard synchronization of the intensity fluctuations between the lasers.
	One roundtrip (7.7\,ns) further, at \textpm 12~ns, weaker revival peaks are barely observable.  
	\textbf{(b)} Simulated intensity dynamics, showing the leader-laggard behavior of the two coupled micropillars. 
	The time axis for pillar 2 has been shifted with respect to pillar 1 by 3.85\,ns, i.e., the optical distance between the two micropillars. 
	This illustrates the delayed correlation of the intensity fluctuations.
}
\end{figure}

Determining the photon auto- and cross-correlation function is also highly interesting in the present case of mutually coupled microlasers to obtain profound insight into the underlying emission dynamics and possible synchronization of intensity fluctuations. In the respective experiment the output intensities of pillar P1 and pillar P2 are cross correlated via SPCM\,1 (single photon counting module) and SPCM\,2 as indicated in Fig.~\ref{fig:Setup-Spectra}(a). Polarization optics are used to flexibly detect photons from any polarization mode of pillars P1 and P2. 
We focus our study on the case where the weak modes are resonantly coupled and show pronounced intensity fluctuations, as the strong modes show only marginal signatures of photon bunching 
and no significant cross correlation peaks when resonantly coupled. 
Noteworthy, this is a typical behavior in delay-coupled micropillar lasers~\cite{Albert.2011}. 
We denote the second-order photon correlation function of the weak modes as $g^{(2)}_{\mathrm{w}_i\mathrm{w}_j}$, giving the auto-correlation for pillar $i$ when $i=j$, and the cross-correlation for $i\neq j$.
An exemplary weak mode-weak mode cross-correlation measurement is shown in Fig.~\ref{fig:Intensity-crosscorrelation-of}(a) for pump currents of $I_{\mathrm{P1}}=27.7\,\mathrm{\upmu A}$ 
and
$I_{\mathrm{P2}}=24.5\,\mathrm{\upmu A}$.
Clear peaks can be observed at $t_2-t_1 \approx 4 \,\mathrm{ns}$, corresponding to the coupling delay of $3.85 \,\mathrm{ns}$ between the microlasers. 
The double-peak structure indicates leader-laggard intensity synchronization of the two micropillars, i.e., if a fluctuation happens in pillar P1, there is a chance that it will be repeated in pillar P2 and vice versa. 
The numerical time series depicted in Fig.~\ref{fig:Intensity-crosscorrelation-of}(b) confirm this interpretation of the experimental data in terms of leader-laggard dynamics \cite{MUL04}, showing a strong similarity between the time-series when either of the time-series is shifted in time by the coupling delay $\tau$. 
The laser coupling can be observed to irregularly induce short mode-switching events in both lasers (e.g., near $t=153$\,ns for pillar P1 in Fig.~\ref{fig:Intensity-crosscorrelation-of}(b)). 
The relatively low peak values of the cross-correlation $g^{(2)}_{\mathrm{w}_1\mathrm{w}_2}(\tau)$ in comparison to the free-running auto-correlation $g_{\mathrm{w}_1\mathrm{w}_1}^{(2)}(0) = 1.5$ for pillar 1 and $g_{\mathrm{w}_2\mathrm{w}_2}^{(2)}(0) = 1.6$ for pillar 2 is proof of an imperfect synchronization between the lasers, and suggests that only a small ratio ($\approx 13$\%) of switching events are repeated in the respective other laser.

\begin{figure}
	\includegraphics{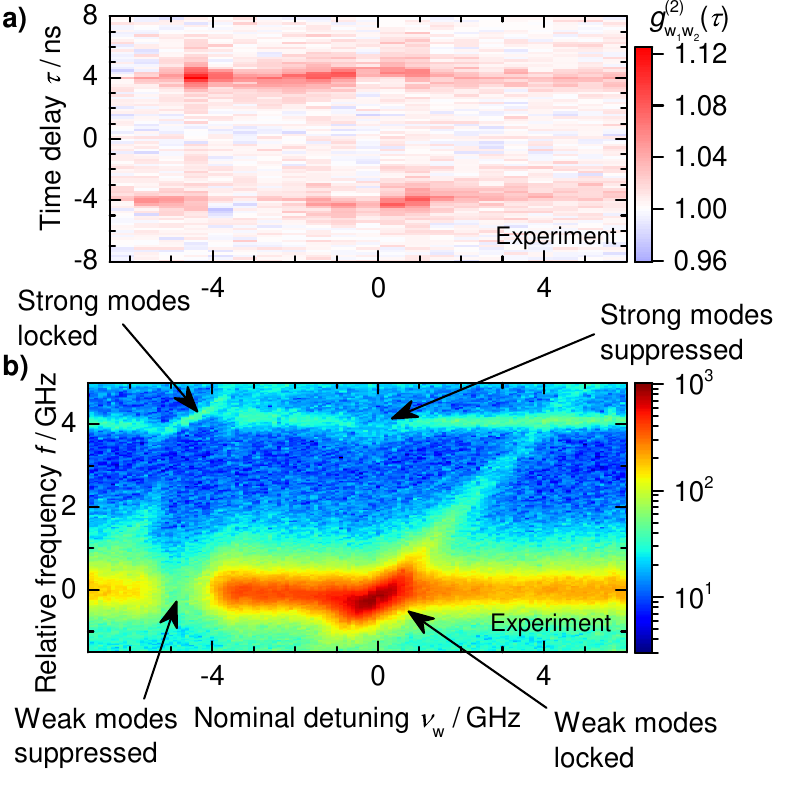}\\
	\vspace{0.2cm}
	\includegraphics{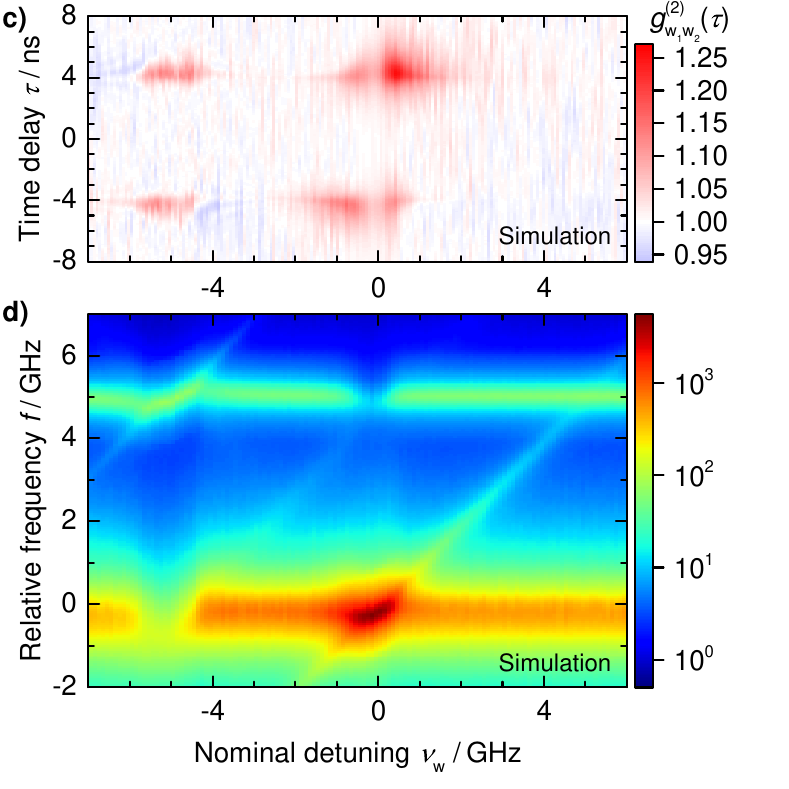}
	\caption{\label{fig:mode-suppression}
%
        Measured (\textbf{a}) and simulated (\textbf{c}) weak-mode intensity cross-correlation $g_{\mathrm{w}_1\mathrm{w}_2}^{(2)}(\tau)$ (color-coded) in dependence of the time delay $\tau$ for different detunings $\nu_\mathrm{w}$. $I_{\mathrm{P1}}=28.0\,\mathrm{\upmu A}$ and $I_{\mathrm{P2}}=25.6\,\mathrm{\upmu A}$ in the experiments.
        \textbf{(b)}, \textbf{(d)} Corresponding log-intensity FPI spectra (color-coded) of the laser output in dependence of the detuning $\nu_\mathrm{w}$. $I_{\mathrm{P1}}=28.6\,\mathrm{\upmu A}$, $I_{\mathrm{P2}}=27.0\,\mathrm{\upmu A}$ in the experiments, $I_{\mathrm{P1}}=27.1\,\mathrm{\upmu A}$, $I_{\mathrm{P2}}=24.4\,\mathrm{\upmu A}$ in the simulations.
        }
\end{figure}

The intensity cross-correlation depends on the dynamical susceptibility of the lasers to a perturbing signal, and thus on their ability to reproduce and synchronize to the signal of the other laser. 
We therefore investigate the dependence of the cross-correlation on the mutual laser detuning $\nu_\mathrm{w}$ of the weak modes. Fig.~\ref{fig:mode-suppression}(a),(b) show the measured cross-correlation of the weak modes of the two lasers and the FPI spectra of the weak mode, respectively.
Since the strong mode is much more intense than the weak mode, it is still visible in the log-intensity-scaled FPI spectra even after attenuation by the polarizing beam splitter. 
The mutual locking of the weak modes around a weak mode detuning of $0$ leads to a strong enhancement of the weak mode signals, while suppressing the strong mode intensity. 
Near the locking range of the strong modes, at a weak mode detuning of $\nu_\mathrm{w}\approx-5$\,GHz, the reverse effect is observed together with a strong suppression of the weak modes. 
This can be understood by the reduction of effective optical losses of the weak mode by 2.5\%, thus reducing the required inversion of the QDs to maintain lasing and reducing the available gain for the strong modes, as known from two-mode lasers in other setups \cite{OSB12,VIR13b,MEI17}. 
In Fig.~\ref{fig:mode-suppression}(c),(d) the corresponding simulated cross-correlation and optical spectra are shown, matching the experimental data very well. In order to reproduce the conditions from Fig.~\ref{fig:mode-suppression}(b), attenuated simulated strong-mode spectra were superimposed onto the simulated weak-mode spectra in Fig.~\ref{fig:mode-suppression}(d). Within the locking range of the weak modes, intensity fluctuations are generally suppressed, thus leading to smaller delay peaks in the cross-correlation. 
At either edge of the locking range, $\nu_\mathrm{w}\approx\pm1.5\text{GHz}$, the signature of the dynamic unlocking of both lasers becomes evident, leading to stronger peaks in the $g^{(2)}_{\mathrm{w}_1\mathrm{w}_2}$ cross-correlation. 
Depending on the detuning, the cross-correlation peak $\pm\tau$ can be enhanced, i.e., the role of the leader in the leader-laggard synchronization of the microlasers is mainly taken on by the laser that is positively frequency-detuned with respect to the other laser.
This asymmetry in the frequency detuning is due to the amplitude-phase coupling, i.e. non-zero $\alpha$~\cite{Ozaki2009}. 
An enhancement of the weak-mode correlations can be observed also within the locking range of the strong modes, as the weak modes are suppressed and driven further towards thermal (bunched) emission. 
For scenarios where strong correlation between the coupled laser emission is required, a detuning near the locking boundaries of the weak modes or within the strong mode locking range should be preferred.

\begin{figure}
	\includegraphics{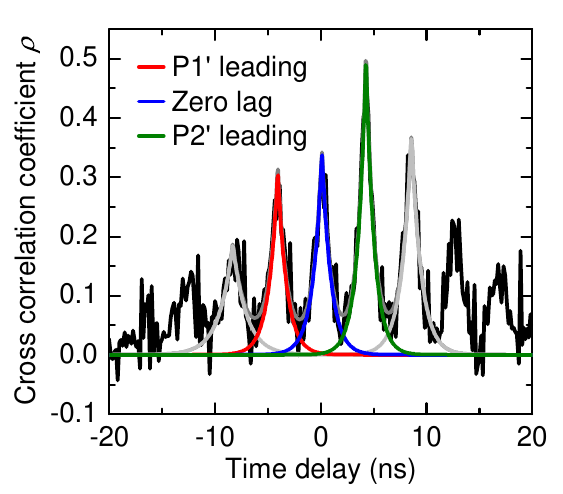}
	\caption{\label{fig:CrosscorrelationCoeff}Delay-dependent intensity linear cross-correlation coefficient $\rho(\tau)$ (Eq.~\eqref{eq:rho}) of coupled micropillar lasers with an additional mirror relay. 
	In the delay range $\left[-10\,\mathrm{ns}, 10\,\mathrm{ns}\right]$ the sum (dark grey line) of five peaks of the form $A \exp(-\left|\tau-\tau_\mathrm{center}\right|/\tau_\mathrm{corr})$  are fitted to the data (black). The zero-lag peak is depicted in blue, the leader-laggard peaks where pillar 1' or pillar 2' is leading are depicted in red and green, respectively.}
\end{figure}

Previous work showed the possibility of zero-lag synchronization of chaotic intensity fluctuations in small networks of mutually coupled semiconductor lasers, in particular if the lasers are also subject to feedback \cite{Aviad.2012}. We explore this important regime of coupled nonlinear oscillators, which could eventually be linked to entanglement of mutually coupled quantum systems in the single photon regime~\cite{Mari2013},  in a setting (see Fig.~\ref{fig:Scheme}(b)) where the mutually coupled micropillars are a subject to self-feedback~\cite{Aviad.2012}.
We therefore explore the possibility of zero-lag synchronization by introducing a mirror relay in the center of the beam path between the two oscillators. The length of the feedback beam path is chosen to introduce additional self-feedback to each cavity-enhanced microlaser with a delay equal to the coupling delay time between the pillars. 
A semipermeable mirror is thus placed at half distance in the coupling path, such that it introduces feedback of the required timing. 
As seen in the previous discussion of Fig.~\ref{fig:mode-suppression}, a strong cross-correlation between the coupled lasers can be expected in regions of dynamical instabilities. We therefore choose two other micropillar lasers, P1' and P2' (see SI for more details), from the same arrays and couple them with a semipermeable mirror in the aforementioned setup. These pillars show a crossing of their strong mode and weak mode intensity in their current dependence at pump currents far above threshold and exhibit more frequent mode-switching events between their respective strong and weak modes \cite{Redlich.2016}. The strong mode competition at this operating point results in a striking increase in the autocorrelation $g_{w_iw_i}^{(2)}(0)$ and an enhanced sensitivity with respect to optical feedback \cite{Holzinger.2018}, which should enhance the correlation signatures when coupling the two microlasers. 
In order to quantify the cross-correlation $g_{w_{1'}w_{2'}}^{(2)}(\tau)$, we calculate the linear intensity cross-correlation coefficient for the two coupled pillars
\begin{align}
\rho(\tau) & = \frac{g^{(2)}_{\mathrm{w_{1'}w_{2'}}}(\tau)-1}{\sqrt{\left(g^{(2)}_{\mathrm{w_{1'}w_{1'}}}(0)-1\right)\left(g^{(2)}_{\mathrm{w_{2'}w_{2'}}}(0)-1\right)}}\text{,}\label{eq:rho}
\end{align}
and expect a value of 1 (-1) for fully linearly correlated (anti correlated) dynamics and a value of 0 for uncorrelated dynamics. 
The resulting time-dependent correlation coefficient is displayed in Fig.~\ref{fig:CrosscorrelationCoeff}. 
With the additional self-feedback due to the semipermeable mirror, additional correlation peaks at a time delay of zero appear (blue line) if compared to Fig.~\ref{fig:Intensity-crosscorrelation-of}(a), along with revival peaks after integer multiples of the coupling delay.  While the cross-correlation measurement shows zero-lag correlation coefficients of up to 34\,\%, a strong peak of up to 50\,\% at the coupling delay time (red and green lines) can be seen, corresponding to simultaneously occurring leader-laggard type synchronization. 
The coexistence of both zero-lag and leader-laggard synchronization peaks in the cross-correlation of the high-beta microlasers with high spontaneous emission noise can be interpreted as a coexistence or stochastic transition between the two types of dynamics. 
In that direction, strong noise is known to perturb coupled lasers away from the synchronization manifold \cite{FLU09}, leading to intermittent desynchronization events known as bubbling. 


\section{Conclusion}
\label{sec:conclusions}
In conclusion, we have explored experimentally and theoretically the synchronization of optical oscillators at the crossroads between classical and quantum physics by mutually coupling cavity-enhanced high-$\upbeta$ microlasers. Due to their bi-modal characteristics and the associated gain competition, the applied electrically driven micropillar lasers are particularly sensitive against external perturbations which facilitates comprehensive studies of synchronization at light powers on the order of only 100 nW. Moreover, the high $\upbeta$-factor of these cavity enhanced microlaser introduces significant spontaneous emission noise which plays an important role in the joint dynamics of the coupled lasers. We have identified synchronization of the mutually coupled microlasers by frequency locking of their emission modes with locking ranges well below 1GHz. Besides the small locking range, the locking also remains imperfect as manifested by pronounced deviations of the locking slopes of both lasers. Interestingly, this behavior, which we attribute to the high spontaneous emission noise in our high-$\upbeta$ microlasers, is in striking contrast to macroscopic coupled laser setups, where the unlocking transition is abrupt. Time-resolved intensity cross-correlation measurements show a noise-induced partial synchronization of the intensity patterns, reaching correlation coefficients of up to 50\%. When coupled with an additional passive relay, we even observe signatures of both zero-lag synchronization which co-exists with leader-laggard type of synchronization. Our numerical simulations based on semi-classical stochastic rate-equations reproduce the experimental results very well. Additionally, they reveal a fine structure in the optical spectra of the locked microlasers comprising several compound laser modes, forming a frequency comb with a broad Gaussian envelope. We interpret this mode structure as a stochastic switching between different compound laser modes being individually locked between the coupled microlasers. As such, our results pave the way for studying the synchronization of optical oscillators in the quantum regime, for which intriguing effects such as quantum synchronization blockade~\cite{Lorch2017} have been predicted and where and boundaries between classical synchronization and quantum entanglement phenomena~\cite{Mari.2013, Galve2017} could be explored experimentally in the future.


\section{Acknowledgements}

The research leading to these results has received funding from the European Research Council (ERC) under the European Union's Seventh Framework (ERC Grant Agreement No. 615613). 
B.L. and K.L. acknowledge support from DFG (Deutsche Forschungsgemeinschaft) within CRC787. 
We kindly thank I. Kanter for interesting and fruitful discussions. 


\bibliographystyle{apsrev4-1}

%

\end{document}